\pdfoutput=1

\documentclass{article}
\usepackage{arxiv}

\usepackage[utf8]{inputenc} 
\usepackage[T1]{fontenc}    
\usepackage{hyperref}       
\usepackage{url}            
\usepackage{booktabs}       
\usepackage{amsfonts}       
\usepackage{nicefrac}       
\usepackage{microtype}      
\usepackage{lipsum}         
\usepackage{graphicx}
\usepackage[numbers]{natbib}
\usepackage{doi}
\usepackage{svg}
\usepackage{graphicx}
\usepackage{booktabs}
\usepackage{subcaption}
\newtheorem{definition}{Definition}
\usepackage{amsmath}
\usepackage{indentfirst}
\usepackage[ruled]{algorithm2e}
\SetKwInput{kwInputs}{Inputs}
\usepackage[section]{placeins}
\usepackage{cleveref}       
\setcitestyle{square}
\usepackage{braket}
\usepackage{setspace}
\DeclareMathOperator{\diag}{diag}

\title{Quantum Fault Trees and Minimal Cut Sets Identification}

\author{Gabriel San Martín Silva \\
        The B. John Garrick Institute for the Risk Sciences,\\
        Center for Reliability Science and Engineering,\\
	Department of Civil and Environmental Engineering,\\
	University of California Los Angeles\\
	Los Angeles, CA\\
	\texttt{gsanmartin@ucla.edu} \\
	\And
        Enrique López Droguett \\
	The B. John Garrick Institute for the Risk Sciences,\\
        Center for Reliability Science and Engineering,\\
	Department of Civil and Environmental Engineering,\\
	University of California Los Angeles\\
	Los Angeles, CA\\
	\texttt{eald@ucla.edu} \\
}

\begin{document}

\newcommand\be{\{BE_i\}_{i=1}^{N_{BE}}}
\newcommand\ie{\{IE_i\}_{i=1}^{N_{IE}}}

\newcommand\kbe{\{\ket{BE_i}\}_{i=1}^{N_{BE}}}
\newcommand\kie{\{\ket{IE_i}\}_{i=1}^{N_{IE}}}
\newcommand\ktop{\{\ket{TOP_i}\}_{i=1}^{N_{BE}}}

\maketitle

\begin{abstract}
Fault Trees represent an essential tool in the reliability and risk assessment of engineering systems. By decomposing the structure of the system into Boolean function, Fault Trees allow the quantitative and qualitative analysis of the system. One of the main important tasks in Fault Tree analysis is the identification of Minimal Cut Sets, defined as groups of components that present the least path of resistance toward a system's failure. Identifying them allows reliability engineers to enhance the reliability and safety of the system, making system failures less likely to occur. However, the minimal cut set identification problem is challenging to solve, due to the exponential growth experienced in the number of feasible configurations as the system's size grows linearly. Over the last few years, quantum computation has been heralded as a promising tool to tackle computational challenges of increased complexity. The reason for this is the promising prospects that the use of quantum effects has for challenging computational tasks. However, its application into Probabilistic Risk Assessment and reliability engineering, and in particular to challenges related to the Fault Tree model, is still uncharted territory. To fill this gap, the objective of the paper is to integrate quantum computation into the Fault Tree Model and present an assessment of their capabilities for the minimal cut set identification problem. To this end, this paper proposes a novel algorithm to encode a fault tree into a quantum computer and to perform the identification of minimal cut sets with increased efficiency via the application of the Quantum Amplitude Amplification protocol. For validation purposes, a series of theoretical and numerical results, the latter obtained using a quantum simulator, are presented in which the proposed algorithm is compared against traditional approaches, such as Monte Carlo sampling.

\end{abstract}


\section{Introduction}

Since their creation in the 1960s \cite{watson1961launch} and their development within the nuclear industry in the 1970s \cite{fussell1976fault}, Fault Trees have been used in almost every industry in which the reliability and operational safety of engineering systems is a concern. Today, Fault Trees represent one of the most well-known and used tools in the reliability and risk assessment of complex engineering systems.

One of the main reasons why Fault Trees are so widely used in engineering practice is their versatility, allowing both the qualitative and quantitative analysis of systems from a principled approach. For example, applications of Fault Trees have found use in the nuclear industry \cite{wang2016fault}, electric industry \cite{volkanovski2009application}, manufacturing industry \cite{cheng2013application}, and even in medical applications \cite{rogith2017using}. Moreover, over the last years, Fault Tree models have been adapted to support new requirements from engineering systems, such as modeling time dependencies with dynamic fault trees \cite{baek2021application, kabir2020hybrid} and fuzzy logical relationships with fuzzy fault trees \cite{mahmood2013fuzzy}. Among the multiple applications of Fault Tree Analysis (FTA), one of the most relevant tasks is the identification of Minimal Cut Sets (MCS) \cite{vatn1992finding}. MCS can be defined as an irreducible set of components whose failure triggers the failure of the whole system. They are irreducible because the failure of all elements that compose an MCS is required for the system’s failure to occur. MCSs are important for a system’s reliability because they represent the path of least resistance toward failure. The number of MCS in a system, and more importantly, how many components are in each MCS are a powerful indication of a system’s reliability. 

However, the identification of MCS is a daunting task in most practical applications. The reason for this is two-fold. Firstly, the number of possible outcomes in a Fault Tree model grows exponentially with the number of components in the system, forbidding any brute-force approach to solve the search problem. Secondly, the search space is usually very complex and unstructured, resembling an unordered search instance. As the demand for larger, more interconnected systems increases, it is expected that the identification of minimal cut sets will become an extremely challenging problem, possibly even surpassing the capabilities of traditional computation.

Quantum computing techniques emerge as a promising alternative to tackle problems of increased complexity and size \cite{nielsen2010quantum}. By harnessing quantum effects such as entanglement and superposition to perform computation, quantum devices are able to represent discrete probability distributions that span over exponentially large event spaces.  This represents an ideal ground for the translation of Probabilistic Risk Assessment (PRA) techniques such as Fault Trees models. However, even though quantum computation has been heralded as a promising alternative to traditional computation for years, its development in PRA has been slower than expected. 

To help close this relevant gap, this paper’s objective is to contribute towards the integration and assessment of quantum computation into PRA approaches by proposing a novel algorithm to efficiently identify MCS in a Fault Tree model. For this, we present the Quantum Fault Tree (QFT) model, a quantum algorithm capable of encoding all the logic and probabilistic relationships that the traditional Fault Tree model represents. Using the QFT model as a basis, we further propose a novel algorithm to identify minimal cut sets that has the potential of being orders of magnitude more efficient than traditional Monte Carlo sampling. The proposed algorithm is based on the well-known Quantum Amplitude Amplification algorithm \cite{brassard2002quantum}, one of the few quantum protocols with theoretical assurances of quantum advantage. To validate the proposed approach, we perform both a theoretical and numerical analysis of the algorithm and compare it against both quantum-based and traditional alternatives.

Our main contributions are three. First, we present a brief, yet complete exposition on the main aspects of quantum computation. For this, a probabilistic approach is preferred, abstracting quantum mechanics from the explanation as much as possible. It is expected that this angle will ease the exploration of quantum computation for researchers and practitioners with experience in the reliability, safety , and risk fields. Second, it is introduced the development of the Quantum Fault Tree model as a quantum extension of the traditional Fault Tree model that allows the application of quantum algorithms within the reliability and PRA context. Finally, it is presented the development of a novel algorithm to perform MCS identification in Fault Trees by using the Quantum Amplitude Amplification algorithm and the QFT model.

The paper is organized as follows. Section \ref{sec:ft} introduces the traditional Fault Tree model in an effort of keeping the paper accessible to both risk and reliability researchers and quantum computing scientists. Section \ref{sec:qc} continues with an introduction to quantum computation and in particular the gate-based and quantum circuit model. Section \ref{sec:qft} presents the QFT model, establishing the capabilities and limitations of the approach. The proposed algorithm is presented in Section \ref{sec:mcs}. A theoretical and numerical validation of the proposed approach is presented in Section \ref{sec:exp}. Finally, Section \ref{sec:concluding} contains our concluding remarks, including future opportunities for research in the area.


\section{Fault Tree Analysis and Traditional Minimal Cut Set Identification} \label{sec:ft}

Originally introduced in 1962 at Bell Telephone Laboratories \cite{watson1961launch}, Fault Tree (FT) models are one of the most used methods in contemporary risk and reliability analysis. In their standard formulation, they are described as directed acyclic graphs (DAGs) with a tree-like structure. The leaves in the tree represent a set of basic events $\{BE_i\}_{i=1}^{N_{BE}}$, modeled as binary random variables indicating whether $BE_i$ occurs $(BE_i=1)$ or not $(BE_i=0)$. The root of the tree is a single node commonly referred to as the TOP event. The nodes connecting the basic events and the TOP event are known as intermediary events, $\{IE_i\}_{i=1}^{N_{IE}}$. While the occurrence of basic events is controlled by a probability distribution, the TOP and intermediary events are modeled as the output of a series of Boolean functions that can take two or more inputs from the set $\be \cup \ie$.

In risk and reliability assessments, FTs are used to model the relationships between the failure of basic components in a system (represented as basic events) and the failure of the system itself (represented as the TOP event). The intermediary events are used to represent sub-systems that aid in the functional representation of the system. In this sense, the “occurrence” of an event shall be interpreted as the failure of a component, sub-system, or system.

While the literature contains multiple variants of FTs models (see \cite{ruijters2015fault} for a comprehensive review), this paper will focus on trees with the following characteristics. First, basic events will be represented as independent Bernoulli random variables, $BE_i \sim Be(p_i)$, where $p_i$ is the probability of failure of the i-th basic event. Second, the Boolean functions that generate the intermediary and TOP events will be taken as either the multi-input AND or OR logical gates. Finally, the analysis will exclude non-coherent FTs. This type of FT includes a NOT gate into the set of permissible Boolean functions, allowing situations where the failure of a basic component may prevent the failure of the system, or where a functioning component may contribute towards the failure of the system. An example of a non-coherent situation is where a pump failure may prevent the flow of fuel to a failing engine, therefore avoiding an explosion. 

\begin{figure}[h]
	\centering
        \includegraphics[]{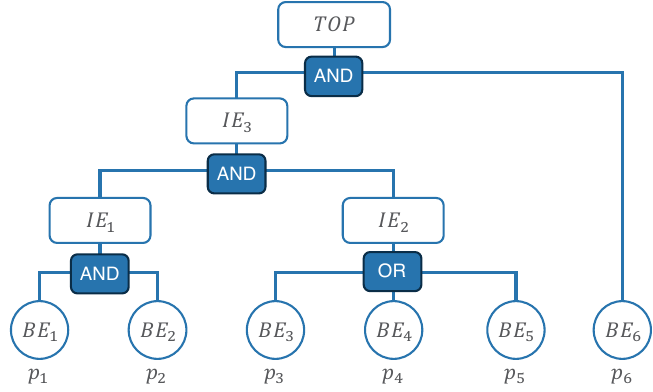}
	\caption{Fault Tree with $N_{BE}=6$  basic events and $N_{IE}=3$ intermediary events.}
	\label{fig:sec2_ft}
\end{figure}

Figure \ref{fig:sec2_ft} depicts an FT composed by $N_{BE}=6$ basic events and $N_{IE}=3$ intermediary events. Note how $IE_3$ receive as inputs the intermediary events $IE_1$ and $IE_2$, while the TOP event takes as inputs a series of intermediary events, in addition to the basic event $BE_6$.

FT models also hold an alternative interpretation that will prove useful for their implementation as quantum-enabled models. Instead of looking at FTs as a tree-like graph, we can model them as the combination of two elements: a probability distribution g over the outcome of the basic events, and a Boolean function $f_{FT}$ that determines the outcome of the TOP event. For this, let us first represent the outcome of the basic events as a binary vector $\vec{x}_{BE} \in \{0,1\}^{N_{BE}}$, where $x_{BE_i}=1$ if the i-th basic event fails, and  $x_{BE_i}=1$ if it does not fail. The distribution $g: \vec{x}_{BE} \rightarrow [0,1]$ assigns a probability mass to every candidate configuration of basic events. If the basic events are modeled as independent Bernoulli random variables, $g$ will be given by the expression shown in Equation \ref{eq:sec_2_1}:

\begin{equation} \label{eq:sec_2_1}
    g(\vec{x}_{BE}) = \prod_{i=1}^{N_{BE}} p_i^{x_{BE_i}}(1-p_i)^{1-x_{BE_i}}
\end{equation}

The second component is the Boolean function $f_{FT}: \vec{x}_{BE} \rightarrow \{0,1\}$, which indicates whether the system fails or not for a certain configuration $\vec{x}_{BE}$. This Boolean function will encapsulate all the relationships given by the logic gates and intermediary events in the FT. With this in mind, we can offer the Fault Tree definition that will be used throughout this paper:

\begin{definition}
  A Fault Tree model is a 3-tuple $FT=\langle \{p_i\}_{i=1}^{N_{BE}}, g, f_{FT}]\rangle$  consisting of a set describing the failure probabilities for the $N_{BE}$ basic events, a probability distribution $g: \vec{x}_{BE} \rightarrow [0,1]$ that assigns a probability to each basic event outcome and a Boolean function $f_{FT}: \vec{x}_{BE} \rightarrow \{0,1\}$ that determines the outcome of the TOP event.
\end{definition}

The analysis of engineering systems using an FT model is referred to as Fault Tree Analysis (FTA) \cite{lee1985fault}. In general, FTA methodologies are grouped into two categories: qualitative and quantitative analysis. While quantitative analysis is concerned with providing numerical metrics of a system’s performance through the determination of its failure probability, qualitative analysis is focused on providing relevant insights using as an input the structure of the tree itself. One of the most important results of qualitative FTA is the identification of cut sets and minimal cut sets (MCS) in a fault tree. A cut set is any configuration of basic events $\vec{x}_{BE}$ that causes the failure of the system,  $f_{FT}(\vec{x}_{BE}) =1$. A MCS is any cut set that is irreducible, i.e., the prevention of any of its basic events’ failures also prevents the failure of the system. The formal definition of an MCS is given below:

\begin{definition}
    A configuration $\vec{x}_{BE}$ is a minimal cut set (MCS) if and only if $f_{FT}(\vec{x}_{BE}) =1 \land f_{FT}(s(\vec{x}_{BE}, i))=0\ \forall i \in F$, where $F \in \{i| x_{BE_i}=1\}$ and $s:\{0,1\}^{N_{BE}} \rightarrow \{0,1\}^{N_{BE}}$ is a "turning-off" function that sets the i-th component of $\vec{x}_{BE}$ to $0$.
\end{definition}

As an example, the FT presented in \ref{fig:sec2_ft} contains three MCS, represented by the sub-sets $\{BE_1,BE_2,BE_3,BE_6\}$, $\{BE_1,BE_2,BE_4,BE_6\}$, and$\{BE_1,BE_2,BE_5,BE_6\}$. In vector notation, these subsets correspond to the configurations $[1,1,1,0,0,1]^T$, $[1,1,0,1,0,1]^T$, and $[1,1,0,0,1,1]^T$, respectively.

MCSs are important in qualitative FTA since they give the “path of least resistance” towards system failure. If an MCS contains only one basic event, then the whole system could fail upon the failure of a single component. Identifying MCS in practical fault trees is computationally expensive due to the exponential increase in candidate configurations given a linear increase in the size of the system. Current approaches to tackle MCS identification include Boolean Manipulation \cite{vesely1981fault}, Binary Decision Diagrams \cite{rauzy1993new}, \cite{coudert1993fault}, and Monte Carlo-based approaches \cite{vesely1970prep}. In this paper, we propose a novel algorithm to identify MCS leveraging the capabilities of quantum computers. To this end, the next section introduces a brief, yet comprehensive introduction to the main concepts of quantum computation. 


\section{Quantum Computation: A Probabilistic Perspective} \label{sec:qc}

For this paper, it suffices to understand a quantum computer as an external computing device capable of performing a small subset of mathematical operations very efficiently. Consequently, their purpose is not to replace traditional computers, but to enhance their capabilities. In this section, we provide a self-contained introduction to quantum computation, focusing on a probabilistic and functional perspective, avoiding topics related to quantum mechanics whenever possible. We start by introducing the \textit{ket} notation, widely used in quantum computing.

\subsection{Ket Notation}

In quantum computation, a complex vector $\vec{\psi}$ is written using the ket notation as $\ket{\psi} \in \mathbb{C}^n$, whereas its conjugate transpose is written using the \textit{bra} notation as  $\vec{\psi}^{\dagger}=\bra{\psi} \in \mathbb{C}^{1\times n}$. The ket notation is also used to represent common operations between vectors and matrices. For this paper, we will make use of three key operations. Firstly, the inner product between two complex vectors, $\ket{\psi}$ and $\ket{\phi}$, is rewritten using the ket notation as $\braket{\psi|\phi}$. Secondly, the Kronecker product between two column vectors is rewritten as $\ket{\psi} \otimes \ket{\phi}$, or by one of their alternative representations $\ket{\psi} \ket{\phi}$ or $\ket{\psi\phi}$. The result of this operation is shown in Equation \ref{eq:sec_3_1}:

\begin{equation}\label{eq:sec_3_1}
    \ket{\psi} \in \mathbb{C}^n, \ket{\phi} \in \mathbb{C}^p \rightarrow \ket{\psi} \otimes \ket{\phi} = \begin{pmatrix} \psi_1 \ket{\phi} \\ \vdots \\ \psi_n \ket{\phi} \end{pmatrix} \in \mathbb{C}^{np}
\end{equation}

Finally, the Kronecker product between a column vector and a row vector is written as $\ket{\psi} \otimes \bra{\phi}$. The result of this operation is shown in Equation \ref{eq:sec_3_2}. Note how in this case the result is a matrix instead of a vector.

\begin{equation}\label{eq:sec_3_2}
    \ket{\psi} \in \mathbb{C}^n, \ket{\phi} \in \mathbb{C}^p \rightarrow \ket{\psi} \otimes \bra{\phi} = \begin{pmatrix} \psi_1 \bra{\phi} \\ \vdots \\ \psi_n \bra{\phi} \end{pmatrix} \in \mathbb{C}^{n\times p}
\end{equation}

We will use this notation to describe, from a mathematical point of view, the set of operations performed by a quantum computer. These operations are the creation, measurement, and modification of the quantum state. We start by reviewing the first operation below.

\subsection{Creation of Quantum States}

Information in a quantum computer is stored in the form of a quantum state, mathematically represented as a complex vector $\ket{\psi} \in \mathbb{C}^n$ with unit length. The smallest quantum state possible is a two-dimensional complex vector known as \textit{qubit} and denoted as $\ket{q} \in \mathbb{C}^2$. Quantum computers have access to a \textit{registry} of $N$ qubits that they can combine to generate larger quantum states. The combination process is described in Equation \ref{eq:sec_3_3}. Note that this is the same expression shown in Equation \ref{eq:sec_3_1}, but expanded to the case of $N$ complex vectors. 

\begin{equation}\label{eq:sec_3_3}
    \ket{\psi} = \bigotimes_{i=0}^{N-1} \ket{q_i} \in \mathbb{C}^{2^N}
\end{equation}

Due to the properties of the Kronecker product, combining the qubits of a registry of length $N$ results in a quantum state of dimension $2^N$. 

A quantum state can be decomposed into the weighted sum of $n=2^N$ linearly independent vectors. For simplicity, quantum computing theory uses the canonical basis $\{\ket{e_i}\}_{i=0}^{2^N-1}$ to represent this weighted sum. The decomposition operation is shown in Equation \ref{eq:sec_3_4}:

\begin{equation}\label{eq:sec_3_4}
    \ket{\psi} = \sum_{i=0}^{2^N-1} c_i\ket{e_i},\ \ \ c_i \in \mathbb{C}
\end{equation}

Qubits can also be written following \ref{eq:sec_3_4} as $\ket{q} = c_0 \ket{0} + c_1 \ket{1}$, where the kets $\ket{0}$ and $\ket{1}$ represent the basis vectors $[1,0]^T$ and $[0,1]^T$, respectively. The kets $\ket{0}$ and $\ket{1}$ will prove to be very useful for interpreting the results of measurement operations in quantum computation. The measurement operation is described in the following section.

\subsection{Measurement of Quantum States}

The most differentiating factor that separates quantum and traditional computers is how information is extracted out of them. Due to restrictions derived from quantum mechanics, it is impossible to retrieve from the quantum device the full set of complex coefficients that compose a quantum state $\ket{\psi}$. Instead, the quantum computer performs a process known as \textit{measurement}. The measurement operation allows the user to sample an element from the set $\{\vec{x}_i\}_{i=0}^{2^N-1}$, i.e., the space of all possible binary vectors of length $N$. The sampling is performed in accordance with a probability distribution. To describe this probability distribution, first note that a bijective map can be established between the set $\{\vec{x}_i\}_{i=0}^{2^N-1}$ and the quantum state’s canonical basis\footnote{In the canonical basis, $\ket{e_i}$ is the i-th column of an identity matrix of size $n$.} $\{\ket{e_i}\}_{i=0}^{2^N-1}$ through Equation \ref{eq:sec_3_5}:

\begin{equation}\label{eq:sec_3_5}
    \ket{e_i} = \bigotimes_{i=0}^{N-1} \ket{x_{i,j}}
\end{equation}

where $\ket{x_{i,j}}=\ket{0}$ if the j-th component of $\vec{x}_i$ is $0$, and $\ket{x_{i,j}}=\ket{1}$ if the j-th component of $\vec{x}_i$ is $1$. A simpler interpretation of this mapping is to think of $\vec{x}$ as a vector containing the binary representation of the integer $i$. Equation \ref{eq:sec_3_5} allows us to describe the probability distribution induced by the measurement operation as $p(\vec{x}_i) = ||c_i||^2$ where $c_i$ is the complex coefficient accompanying the canonical basis vector $\ket{e_i}$ given by Equation \ref{eq:sec_3_5}. This distribution can also be represented using the dot product formula, as shown in Equation \ref{eq:sec_3_6}:

\begin{equation}\label{eq:sec_3_6}
    p(\vec{x}_i) = \braket{\psi | \otimes_{j=0}^{N-1} \ket{x_j}}^2 = \braket{\psi | e_i}^2
\end{equation}

Equation \ref{eq:sec_3_6} describes two fundamental features of quantum computation. Firstly, we can interpret a quantum state as a mathematical object encoding a discrete probability distribution over the space of binary vectors of length $N$. Secondly, Equation \ref{eq:sec_3_6} justifies why quantum states are required to have a unit length: for $p$ to be a valid probability distribution, the sum of the squared complex amplitudes needs to add up to one. 

In the case of a single qubit, the measurement operation allows us to interpret the quantum state $\ket{q}$ as a binary random variable. To see this, let us first express $\ket{q}$ as $\ket{q}=c_0\ket{0} + c_1 \ket{1}$. Following Equation \ref{eq:sec_3_6}, the probability of obtaining as measurement outcome $x=0$ or $x=1$  is $||c_0||^2$ or $||c_1||^2$, respectively. It follows that by interpreting either of them as a “trial success”, a qubit element is equivalent to a Bernoulli random variable. 

In multi-qubit systems, measurement outcomes also have a very useful interpretation as the results of measuring each individual qubit in the system. For this, we shall identify the outcome of qubit $\ket{q_j}$ with component $x_{i,j}$ of measurement $\Vec{x}_i$. This enables the interpretation of quantum states as stochastic systems composed of $N$ binary random variables. In the following section, we will see that quantum computation allows us to establish rich relationships between these variables, enabling us to encode logical or probabilistic dependencies between them. This is the foundation for formulating the quantum equivalent of traditional models such as Fault Trees. In the following section, we review how these relationships are formed through the transformation of the quantum state.

\subsection{Manipulation of Quantum States} \label{subsec:U}

A quantum computer manipulates the information contained in a quantum state by multiplying the corresponding complex vector with a set of unitary matrices . Equation \ref{eq:sec_3_7} shows the transformation of $\ket{\psi}$ towards $\ket{\psi'}$, through the use of $m$ unitary operators:

\begin{equation} \label{eq:sec_3_7}
    \ket{\psi'} = \prod_{j=1}^{m} U_j\ket{\psi},\ \ \ U_j \in \mathbb{C}^{2^N \times 2^N}
\end{equation}

The act of defining the set of unitary matrices is referred to as \textit{quantum algorithmic design} in the field of quantum computation. In a similar fashion to how $N$ qubits can be combined to form a quantum state, a unitary matrix $U_j$ can also be defined as the combination of $N_{U_j}\leq N$ unitary matrices of smaller dimensions that are applied over one or more qubits in parallel, as shown in Equation \ref{eq:sec_3_8}:

\begin{equation} \label{eq:sec_3_8}
    U_j= \bigotimes_{i=1}^{N_{U_j}} U_{j,i}
\end{equation}

The set of smaller unitary matrices that compose $U_j$ are commonly referred to as \textit{quantum gates}. The only condition for decomposing a unitary operation into a set of quantum gates is that the dimensionality of the resulting matrix must match the dimensionality of the quantum state, in order to maintain the operation valid. This is achieved when quantum gates are applied over all qubits. If for a certain stage $j$, the quantum algorithm does not require a qubit to be operated with a quantum gate, the $2\times 2$ identity matrix can be used to \textit{pad} these qubits and maintain the operation dimensionally coherent. 

Quantum algorithms are usually represented using a wire-shaped diagram known as a \textit{quantum circuit}. Quantum circuits represent, from left to right, all the operations that are applied to an $N$ qubit registry during its transformation process. A quantum circuit diagram must represent three key elements. The first element, located at the left of the diagram, is the quantum registry. The quantum registry indicates how many qubits are to be used in a particular computation and what is the initial quantum state in the system. Unless a different initialization is specified, it is often assumed that all qubits in the registry start in the $\ket{0}$ state. The second element, located in the middle, is the set of quantum gates that will be applied to each qubit or group of qubits to generate the unitary matrices $\{U_j\}$. Finally, the last element of the quantum circuit is the measurement operation, located at the right, which indicates that Equation \ref{eq:sec_3_6} should be used to obtain a stochastic output out of the quantum algorithm. 

\begin{figure}[h]
	\centering
        \includegraphics[]{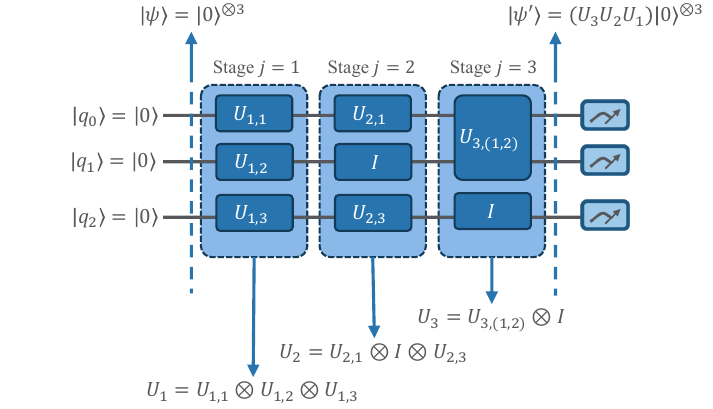}
	\caption{Graphical depiction of the transformation process of a quantum state, also known as a \textit{quantum circuit}. Note how an initial state, $\ket{\psi}$, formed by the combination of three qubits initialized as $\ket{0}$, is subjected to a series of quantum gates. The first operation, $U_1$, is composed by the application of single-qubit gates applied over all the qubits in the registry. The second operation, $U_2$, only receives single qubit operations in the first and third qubits, and therefore the second qubits need to be padded with an identity operation to maintain that stage dimensionally coherent. Finally, the operation $U_3$ applies a double qubit gate over the first and second qubit, while padding the third one with a unitary operation. The global unitary in this case can be computed by the matrix multiplication of the resulting stage unitary gates.}
	\label{fig:sec3_qc}
\end{figure}

While in principle any unitary matrix can be used as a quantum gate to transform a quantum state, quantum algorithms are often designed from a reduced set of quantum gates that are easily implemented in a quantum device. In what follows, we focus our presentation on those quantum gates that are useful for the purposes of this paper. For a complete overview of quantum gates, the reader is referred to \cite{nielsen2010quantum}.

The first quantum gate we will review is the Pauli-X gate, $\sigma_X$. Its definition is presented in Table \ref{tab:qg}. Note that its dimension is $2\times2$, and therefore it is applied over a single qubit gate. The effect of the Pauli-X gate is to simply switch the complex coefficient of the qubit over which it operates, i.e., if $\ket{q} = c_0\ket{0}+c_1\ket{1}$, then $\sigma_X\ket{q} = c_1\ket{0}+c_0\ket{1}$. In particular, it has the important effect of “flipping” the state of a qubit from the $\ket{0}$ to $\ket{1}$, and vice versa. 

The Pauli-Z gate, shown in Table \ref{tab:qg}, is another single-qubit gate. The effect of the Pauli-Z gate is to transform $\ket{q} = c_0\ket{0}+c_1\ket{1}$ into $\ket{q} = c_0\ket{0}-c_1\ket{1}$. More importantly, the effect of a Pauli-Z gate applied over the i-th qubit in an $N$ qubit registry is to apply a negative sign to all states in which the i-th qubit would be measured as $\ket{1}$. This feature is fundamental for the implementation of \textit{oracle operations} in the Quantum Amplitude Amplification algorithm, as explained in Section \ref{subsec:qaa}.

The Hadamard gate, also shown in Table \ref{tab:qg}, is usually applied to transform the state $\ket{q}=\ket{0}$ into $\ket{q}=(1/\sqrt{2}) (\ket{0}+\ket{1})$. When the Hadamard gate is applied to every qubit in a registry of length $N$, it has the interesting property of converting the initial state $\ket{\psi}=\ket{0}^{\otimes N}$ into $\ket{\psi} = (1/\sqrt{2^N}) \sum_{i=0}^{2^N-1}\ket{e_i}$, i.e., the underlying probability distribution encoded in the quantum state is uniform across all potential outcomes. 

The final single-qubit gate that we will review is the Rotation-Y (RY) gate. Contrary to the Pauli and Hadamard gates, the RY gate incorporates a free parameter $(\theta)$  in its definition, and for that reason it is commonly used to integrate external information into quantum circuits. We will use it in the implementation of the Quantum Fault Tree model. The definition of the RY gate is presented in Table \ref{tab:qg}. When applied over a single qubit, the effect of the RY gate is to modify its probability of measuring $\ket{0}$ or $\ket{1}$. In particular, when the qubit is prepared in the $\ket{q}=\ket{0}$ state, the RY gate produces the state $\ket{q'}=RY(\theta)\ket{q}=\cos(\theta/2)\ket{0}-\sin(\theta/2)\ket{1}$, i.e., the probability of measuring $\ket{0}$  and $\ket{1}$ are fully parameterized by the angle $\theta$. 

Quantum gates can also be applied over multiple qubits. The most notorious example is the Control-NOT (CNOT) gate, also defined in Table \ref{tab:qg}. The effect of the CNOT gate is to switch the third and fourth complex coefficients of a quantum state formed by the combination of two qubits, i.e., $\ket{\psi}\in \mathbb{C}^4$. This transformation has a profound practical implication: it generates a conditional dependence between the outputs of both qubits, applying a NOT gate on the measurement of the second qubit if the first qubit is measured as $\ket{1}$. For this reason, the first and second qubits are commonly referred to as “control” and “target” qubits in the context of the CNOT gate. This behavior can be extended to a registry of $N$ qubits, with the first $N-1$ qubits acting as “controls”, and the last one acting as the “target” qubit. The target qubit is only flipped through the action of a NOT gate if all the controls are measured as $\ket{1}$. Note that this “multi CNOT” gate is the quantum equivalent of an AND gate over $N-1$  inputs. We will use it extensively for the quantum encoding process of fault trees, described in Section \ref{sec:qft}.

A fundamental feature of quantum computation is that quantum gates are by definition, reversible. The practical effect of this is that any computation stage in the algorithm can be reversed to obtain a previous quantum state by applying its conjugate transpose, i.e., 
$\ket{\psi} = UU^{\dagger}\ket{\psi}$. We will make extensive use of this property in the proposed MCS identification approach to reduce the number of quantum resources (qubits) required in the algorithm.

\begin{table}[h]
\centering
\caption{Quantum Gates used in this paper.}
\label{tab:qg}
\begin{tabular}{lllll}\toprule
Quantum Gate & Symbol       & Matrix Expression & \# of Qubits & Parametric?     \\ \midrule \addlinespace[1.5ex]
Pauli-X      & $\sigma_X$   & $\begin{bmatrix} 0 & 1 \\ 1 & 1  \end{bmatrix}$      & 2      & No.     \\ \addlinespace[1.5ex] \midrule \addlinespace[1.5ex]
Pauli-Z      & $\sigma_Z$   &  $\begin{bmatrix} 1 & 0 \\ 0 & -1  \end{bmatrix}$     & 2     & No.     \\ \addlinespace[1.5ex] \midrule \addlinespace[1.5ex]
Hadamard     & $H$          &  $\frac{1}{\sqrt{2}}\begin{bmatrix} 1 & 1 \\ 1 & -1  \end{bmatrix}$    & 2            & No.  \\ \addlinespace[1.5ex] \midrule \addlinespace[1.5ex]
Rotation-Y   & $RY(\theta)$ &  $\begin{bmatrix} \cos{\theta/2} & -\sin{\theta/2} \\ \sin{\theta/2} & \cos{\theta/2}  \end{bmatrix}$     & 2   & Yes, with parameter $\theta$. \\ \addlinespace[1.5ex] \midrule \addlinespace[1.5ex]
Control-NOT  & $CNOT$       &   $ \begin{bmatrix} 1 & 0 & 0 & 0 \\ 0 & 1 & 0 & 0 \\ 0 & 0 & 0 & 1 \\ 0 & 0 & 1 & 0  \end{bmatrix}  $                & 4            & No.                           \\ \addlinespace[1.5ex]
\bottomrule
\end{tabular}
\end{table}


\section{Quantum Fault Tree Model} \label{sec:qft}

In this section, we present an algorithmic methodology to translate a fault tree model into a quantum circuit. This approach constitutes a formalization of our work previously presented in \cite{silva2022quantum}. Our scope considers fault trees composed of a set of $\be$ basic events, $\ie$ intermediary events, and one TOP event. We assume that the basic events can be modeled as independent Bernoulli random variables, with probabilities of failure $\{p_i\}_{i=1}^{N_BE}$. Additionally, it is assumed that the intermediary event $i$ can be described as the result of applying the logic operator $O_i \in \{AND,OR\}$ to the set of inputs $S_i, i\in\{1,...,N_{IE}\}$. In a similar fashion, the TOP event is assumed to be generated by applying the logic operator $O_{TOP}$ to the set of inputs $S_{TOP}$. Note that for both the intermediary and TOP events, the input set could include both intermediary and basic events.

The encoding algorithm is divided into four successive stages. The first stage is the preparation stage, where  three-qubit registries, $\kbe$, $\kie$, and $\{\ket{TOP}\}$, are created. As the symbology indicates, these qubit registries will store the outcomes of the basic events, intermediary events, and TOP events, respectively. We consider the measurement outcomes $\ket{0}$ and $\ket{1}$ to represent an operational and failed status, respectively. All qubits in all registries are initialized in the $\ket{0}$ state.

The second stage in the algorithm involves encoding the probability distribution generated by the basic events into the basic event registry. Since each basic event is an independent Bernoulli random variable, their joint probability distribution is given by Equation \ref{eq:sec_2_1}, which is copied here in Equation \ref{eq:sec_4_1} for the reader’s convenience.  In Equation \ref{eq:sec_4_1}, $\vec{x}_{BE}\in \{0,1\}^{N_{BE}}$ is a binary vector that indicates whether basic event $i$ has failed $(x_{BE_i}=1)$, or not $(x_{BE_i}=0)$. 

\begin{equation} \label{eq:sec_4_1}
    g(\vec{x}_{BE}) = \prod_{i=1}^{N_{BE}} p_i^{x_{BE_i}}(1-p_i)^{1-x_{BE_i}}
\end{equation}

As mentioned in Section \ref{subsec:U},  $g(\vec{x}_{BE})$ can be encoded by applying a $RY(\theta_i)$ quantum gate to each qubit $\ket{BE_i}$ in the registry. The rotational angle $\theta_i$ depends on the correspondent basic event failure probability and is given by Equation \ref{eq:sec_4_2}. The resulting unitary operation will be denoted as $U_{BE}$, since it encodes into the quantum circuit the probability distribution over the basic events outcomes.

\begin{equation} \label{eq:sec_4_2}
    \theta_i=2 \mathrm{atan} \left( \sqrt{\frac{p_i}{1-p_i}} \right)
\end{equation}

The third stage in the algorithm corresponds to the translation of the intermediary events into the correspondent registry. For this, the approach to encode $IE_i$ depends on the correspondent logical operator $O_i$. If $O_i=AND$, then $IE_i$ can be encoded into the quantum circuit by applying a multi-control NOT (MCNOT) gate using as inputs the qubits contained in the set $S_i$, and as output the qubit $\ket{IE_i}$. To see why this works, remember that the MCNOT gate flips the output qubit if and only if the input qubits are all measured in the $\ket{1}$ state; otherwise, it is left unmodified. This results in a behavior that matches exactly the logical AND gate: if any of the inputs is measured as $\ket{0}$, then the output qubit will be left unmodified and therefore measured as $\ket{0}$ (its initial state). On the other hand, if all inputs are measured as $\ket{1}$, then the  MCNOT gate will flip the output qubit from $\ket{0}$ to $\ket{1}$, returning the expected result.

If $O_i=OR$, then $IE_i$ can be encoded into the quantum circuit by first transforming the OR logical gate into an equivalent representation using Equation \ref{eq:sec_4_3}, where the symbol $\neg$ indicates a NOT operation.

\begin{equation} \label{eq:sec_4_3}
    OR(s_1,s_2,s_3,...) = \neg AND(\neg s_1,\neg s_2,\neg s_3,...)
\end{equation}

Equation \ref{eq:sec_4_3} shows that the result of an OR gate over the input set $S$ is equivalent to first applying an AND gate over the set  $\neg S$, and then applying a NOT gate to the result. With this relationship, the logical OR gate can be implemented in a quantum circuit by first applying a series of Pauli-X gates to all input qubits in $S$, followed by an MCNOT gate and finally a Pauli-X gate over the output qubit $\ket{IE_i}$. The result is a quantum operation that exactly reflects the behavior of the logical OR gate. However, in a quantum circuit, we finalize the operation by applying an additional layer of Pauli-X gates over the input qubits contained in $S$, in order to return them to their original states.

The encoding process described above can be iterated for each intermediary event in the fault tree. The only ordering consideration lies in ensuring that if intermediary event $IE_i$ is utilized as input for intermediary event $IE_j$, then the logic gate that generates $IE_i$ should be encoded before the logic gate that generates $IE_j$. The resulting operation from this stage will be denoted as $U_{IE}$, since it encodes the logic dependencies required to compute the outcomes of the intermediary events.

The fourth stage of the algorithm is the encoding of the TOP event. For this, note that the TOP event can be treated as another intermediary event, and therefore the same set of rules described for the intermediary events can be applied in its encoding process. We denote this unitary operation as $U_{TOP}$.

The result of executing and measuring the resulting quantum circuit is a binary vector  $\vec{x} \in \{0,1\}^N$, where $N=N_{BE}+N_{IE}+1$. This vector can be decomposed into two vectors and one scalar:  $\vec{x}_{BE}$, $\vec{x}_{IE}$, and $x_{TOP}$, of dimension $N_{BE}$, $N_{IE}$, and one, respectively. This decomposition is structured such that $\vec{x} = [\vec{x}_{BE}, \ \vec{x}_{IE},\ x_{TOP}]$. The vector $\vec{x}_{BE}$ contains the final state of the basic events, distributed in accordance with $g(\vec{x}_{BE})$. The vector $\vec{x}_{IE}$ contains the final state of the intermediary events and will always fulfill the logical relationships imposed by the corresponding logical operators. Finally, the scalar $x_{TOP}$ represent the state of the TOP event, fulfilling its logical relationships with elements in $\vec{x}_{BE}$ and $\vec{x}_{IE}$. In other words, both $\vec{x}_{IE}$ and $x_{TOP}$ are deterministic functions of $\vec{x}_{BE}$.

Since the resulting unitary operation encodes the complete behavior of the original fault tree, we will refer to it from here onwards as $U_{FT}$. The operation $U_{FT}$ can also be computed from its individual components as $U_{FT}=U_{TOP}U_{IE}U_{BE}$. The quantum state that results from applying this operation to a qubit registry initialized in the $\ket{0}^{\otimes N}$ state will be referred to as a “Quantum Fault Tree” (QFT). A characteristic of QFT is that the results obtained by executing and measuring them are equivalent to those obtained by first performing Monte Carlo sampling in the basic events random variables, and then propagating their results through the Boolean structure of the fault tree. Note that, as it will be discussed in Section \ref{subsec:mcs}, creating a quantum representation of a fault tree allows us to leverage other quantum computing protocols to greatly enhance aspects of fault tree analysis. In our case, we will couple the QFT model with the Grover search algorithm to propose an efficient minimal cut set identification algorithm. 

Algorithm \ref{Algo1} summarizes the proposed framework to encode a fault tree model into a quantum circuit.

\begin{algorithm}[h]
\kwInputs{\\ \begin{itemize}
 \item Set of basic events $\be$, where each basic event is associated with a probability of failure $p_i$.
 \item Set of intermediary events $\ie$, where each intermediary event is associated with a generating logical operator $O_i\in\{AND,OR\}$, and a set of inputs $S_i$. The set of intermediary events is ordered in such a way that $IE_i$ is not an input of $IE_j$ if $i\leq j$.
 \item A TOP event, described by a generating logical operator $O_{TOP} \in \{AND,OR\}$, and a set of inputs $S_{TOP}$.
\end{itemize}}

\KwResult{A unitary operation $U_{FT}$ that encodes a Fault Tree into a quantum circuit, generating a Quantum Fault Tree (QFT) model.}
\textbf{Algorithm:}
\begin{enumerate}
\item Preparation of qubit registries:
        \begin{itemize}
            \item Initialize a qubit registry composed of the set of qubits $\kbe$ initialized in the $\ket{0}^{\otimes N_{BE}}$ state. 
            \item 	Initialize a qubit registry composed of the set of qubits $\kie$ initialized in the $\ket{0}^{\otimes N_{IE}}$ state. 
            \item 	Initialize a qubit registry composed of qubit $\ket{TOP}$ initialized in the $\ket{0}$ state. 
        \end{itemize}
\item For $i\in \{1,...,N_{BE}\}$:\ \ \ \tcp{Generation of $U_{BE}$}
        \begin{itemize}
            \item 	$\theta_i=2 \mathrm{atan} \left( \sqrt{\frac{p_i}{1-p_i}} \right)$.
            \item 	Apply a $RY(\theta_i)$ quantum gate to qubit $\ket{BE_i}$.
        \end{itemize}

\item For $i\in \{1,...,N_{IE}\}$:\ \ \ \tcp{Generation of $U_{IE}$}
        \begin{itemize}
            \item If $O_i = AND$:
            \begin{itemize}
                \item 	Apply a MCNOT gate using as inputs the qubits in the set $S_i$ and as output the qubit $\ket{IE_i}$.
            \end{itemize}
            \item If $O_i = OR$:
            \begin{itemize}
                \item 	Apply a Pauli-X gate to all input qubits in the set $S_i$.
                \item 	Apply a MCNOT gate using as inputs the qubits in the set $S_i$ and as output the qubit $\ket{IE_i}$.
                \item 	Apply a Pauli-X gate to output qubit $\ket{IE_i}$.
                \item   Apply a Pauli-X gate to all input qubits in the set $S_i$.            \end{itemize}
        \end{itemize}
\item For the TOP event:\ \ \ \tcp{Generation of $U_{TOP}$}
        \begin{itemize}
            \item If $O_{TOP} = AND$:
            \begin{itemize}
                \item 	Apply a MCNOT gate using as inputs the qubits in the set $S_{TOP}$ and as output the qubit $\ket{TOP}$.
            \end{itemize}
            \item If $O_{TOP} = OR$:
            \begin{itemize}
                \item 	Apply a Pauli-X gate to all input qubits in the set $S_{TOP}$.
                \item 	Apply a MCNOT gate using as inputs the qubits in the set $S_{TOP}$ and as output the qubit $\ket{TOP}$.
                \item 	Apply a Pauli-X gate to output qubit $\ket{TOP}$.
                \item   Apply a Pauli-X gate to all input qubits in the set $S_{TOP}$.            \end{itemize}
        \end{itemize}

\item Generate $U_{FT}=U_{TOP}U_{IE}U_{BE}$
\end{enumerate}
 \caption{Quantum Fault Tree encoding approach.}
 \label{Algo1}
\end{algorithm}

It is important to note that the scaling of this encoding algorithm is linear in the number of basic events and intermediary events used in the fault tree. This is a significant improvement over approaches such as the one presented in [20], which uses an approach based on first converting the fault tree model into a Boolean formula described as a clausal normal form (CNF) \cite{brand_d6_2022}, and then encoding each clause separately.

\begin{figure}[h]
\centering
\begin{subfigure}{0.48\textwidth}
    \includegraphics[width=\textwidth]{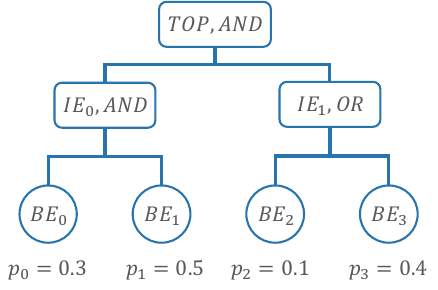}
    \caption{Fault Tree with failure probabilities.}
    \label{fig:first}
\end{subfigure}
\hfill
\begin{subfigure}{0.48\textwidth}
    \includegraphics[width=\textwidth]{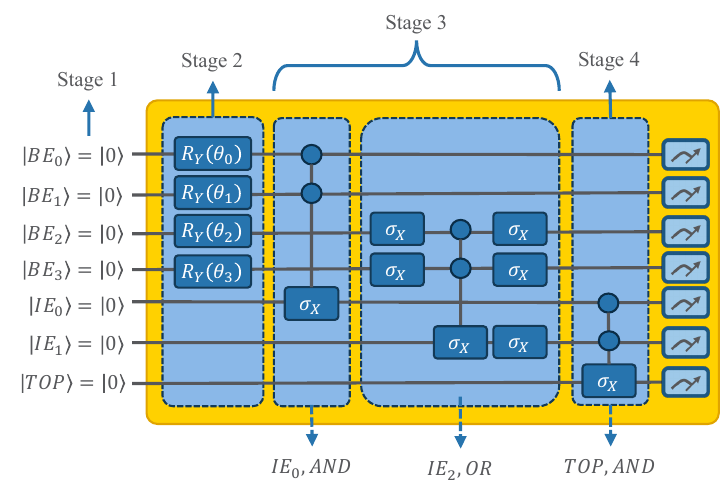}
    \caption{Quantum Fault Tree (QFT) model encoding the Fault Tree.}
    \label{fig:second}
\end{subfigure}
\hfill
\begin{subfigure}{0.9\textwidth}
    \includegraphics[width=\textwidth]{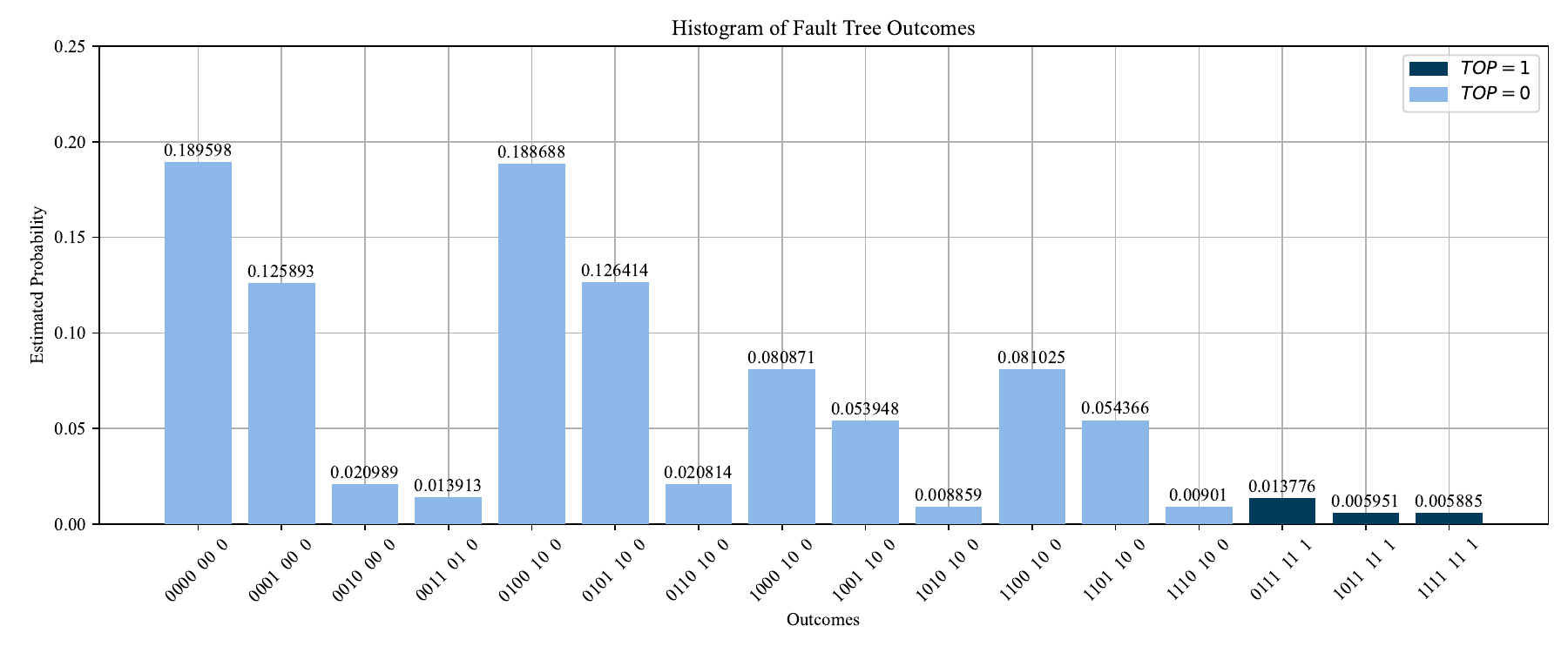}
    \caption{Estimated outcome probabilities from measuring and executing the QFT model $1E6$ times.}
    \label{fig:third}
\end{subfigure}
        
\caption{Example of the QFT framework. (a) Fault Tree composed of $N_{BE}=4$ basic events and $N_{IE}=2$ intermediary events. Failure probabilities are indicated below each basic event. (b) Quantum circuit that results from applying Algorithm \ref{Algo1} to the Fault tree in Figure \ref{fig:sec4} (a). (c) Histogram showing the estimated probabilities for each possible outcome of the fault tree, computed by measuring the quantum circuit a total of $10^6$ times. The first $4$ bits represent the outcomes of the basic events. The fifth and sixth bits represent the outcomes of the intermediary events. The last bit represents the outcome of the TOP event.}

\label{fig:sec4}
\end{figure}

Figure \ref{fig:sec4} shows an illustrative example of the QFT framework, as summarized in Algorithm \ref{Algo1}. For this, a simple fault tree composed of four basic events and two intermediary events is portrayed in Figure \ref{fig:sec4} (a). The fault tree model is encoded into the quantum circuit shown in Figure \ref{fig:sec4} (b), forming a QFT. Note how all stages in the algorithm are represented from left to right in the quantum circuit. Figure \ref{fig:sec4} (c) depicts the estimated probabilities that result from measuring the quantum circuit a total of $1E6$times. Observe how even when seven qubits are used in the quantum circuit, and consequently a total of $2^7=128$ binary vectors could be sampled, in practice we only obtain $2^{N_{BE}}=16$ combinations. The reason for this is that the outcomes of the qubits representing the intermediary and TOP events are not stochastic. Instead, they are a deterministic function of the basic events’ measurement outcomes.

Figure \ref{fig:sec4} (c) also indicates that the proposed algorithm correctly encodes a fault tree into a quantum circuit, including its logical relationship and the underlying probability distribution over basic events. Indeed, the reader can easily verify that the probabilities obtained with the QFT model for each outcome match the theoretical probabilities given by Equation \ref{eq:sec_4_1} and that the logical relationships between events are correctly represented.

\section{Quantum-Enhanced Minimal Cut Set Identification} \label{sec:mcs}

The concepts presented in Sections \ref{sec:qc} and \ref{sec:qft} are used in this section on the novel algorithm to identify minimal cut sets (MCS) in a fault tree by using quantum computation. For this, Section \ref{subsec:qaa} first describes the Quantum Amplitude Amplification (QAA) algorithm, the core component of our approach. Then, Section \ref{subsec:naive} applies the QAA algorithm to an unmodified Quantum Fault Tree (QFT) model to produce what we have termed a \textit{naive MCS identification approach}, so to exemplify the main idea surrounding our proposed technique. Finally, Section \ref{subsec:mcs} modifies the QFT circuit, enabling the automatic identification of MCS, and allowing us to present the algorithm that represents the main contribution in this paper.

\subsection{Quantum Amplitude Amplification Algorithm} \label{subsec:qaa}

The algorithm proposed in this paper to identify MCS in a fault tree uses the Quantum Amplitude Amplification (QAA) algorithm \cite{brassard2002quantum} as a core component. Based on a prior development proposed by Lov K. Grover in 1996 \cite{grover1996fast}, the QAA increases the likelihood of obtaining “marked” outcomes when measuring a quantum state. In this section, we focus our attention on presenting the mathematical formulation of the QAA algorithm from a general perspective. In Section \ref{subsec:naive}, we will connect this framework to the problem of MCS identification in a fault tree, which represents one of the contributions of this paper.

Let us begin the exposition of QAA by describing the setting upon which it is applied. Let us first consider the quantum state $\ket{\psi} = A\ket{0}^{\otimes N}$, produced by multiplying the unitary operation $A\in \mathbb{C}^{2^N \times 2^N}$ with the standard initial state of an $N$ qubit registry. Additionally, let us consider the set of all binary vectors of length $N$, $E=\{\vec{x}_i\}_{i=0}^{2^N-1}$, ordered in such a way that $\vec{x}_i$ is equal to the binary representation of integer $i$. We will assume access to a Boolean function $f:E \rightarrow \{0,1\}$ that divides the elements of $E$ into the subsets $E_1=\{\vec{x}_i \in E| f(\vec{x_i})=1)\}$ and $E_0=\{\vec{x}_i \in E| f(\vec{x_i})=0)\}$. Since the function $f$ separates the elements of $E$ into two distinct sub-sets, we will refer to it as the \textit{oracle} function. It is trivial to verify that $E=E_1 \cup E_0$ and $E_1 \cap E_0 \ \empty$. The oracle function $f$ can be used to generate an oracle unitary operation, $S_f$, shown in Equation \ref{eq:sec_5_1}. The effect of applying $S_f$ over the quantum state $\ket{\psi}$ is to change the sign of all complex coefficients associated with measurement outcomes for which the oracle function returns $1$.

\begin{equation} \label{eq:sec_5_1}
    S_f = \diag\left(\left[ -1^{f(\vec{x}_0)}, -1^{f(\vec{x}_1)}, \dots, -1^{f(\vec{x}_{2^N-1})}\right]\right) \in \mathbb{C}^{2^N \times 2^N}
\end{equation}

Now, let us consider the problem of obtaining elements from $E_1$ as the outcome of measuring $\ket{\psi}$. The probability of measuring an element $\vec{x}_i \in E_1$ is clearly $\sum_{INT(E_1)}||c_i||^2$, where $INT(E_1)$ is a set that contains the integer representation of vectors contained in $E_1$. We will denote this probability with the symbol $a$ for convenience. Consequently, the expected number of measurements required until we find an element from $E_1$ is given by $1/a$. The QAA algorithm allows us to decrease the complexity of this operation to $O(1/\sqrt{a})$, obtaining a quadratic improvement in runtime for this search task.  

To achieve a quantum advantage, the QAA proposes the repeated application of the \textit{Grover} operator $Q$ over the quantum state. The Grover operator is given by $Q=A^{\dagger}S_0AS_f$, where $S_0$ is a unitary matrix that changes the sign of the first complex coefficient in the quantum state, maintaining all the rest untouched. The matrix form of $S_0$ is described in Equation \ref{eq:sec_5_2}:

\begin{equation} \label{eq:sec_5_2}
    S_0 = \diag\left(\left[ -1,0,0,\dots,0 \right]\right) \in \mathbb{C}^{2^N \times 2^N}
\end{equation}

The quantum state that results from applying the Grover operator a total of $j$ times is given by $\ket{\psi'} = Q^j |\ket{\psi}$\$. It can be demonstrated that the resulting quantum state, when measured, has a probability of returning $\vec{x}_i \in E_1$ given by Equation \ref{eq:sec_5_3}. The complete derivation of Equation \ref{eq:sec_5_3} is omitted in this paper for brevity’s sake. However, the interested reader can find the derivation in \cite{brassard2002quantum}:

\begin{equation} \label{eq:sec_5_3}
    p(E_1) = \sin^2((2j+1)\theta_a)
\end{equation}

where $\theta_a = \mathrm{asin}(\sqrt{a})$ In practical terms, Equation \ref{eq:sec_5_3} implies that the application of the Grover operator over $\ket{\psi}$ alters the probabilities of obtaining elements belonging to $E_1$. Note that if the Grover operator is not applied (case $j=0$), then Equation \ref{eq:sec_5_3} reduces to $p(E_1) = \sin^2(\mathrm{asin}(\sqrt{a})) = a$, as expected.  

\begin{figure}[h]
	\centering
        \includegraphics[width=0.95\textwidth]{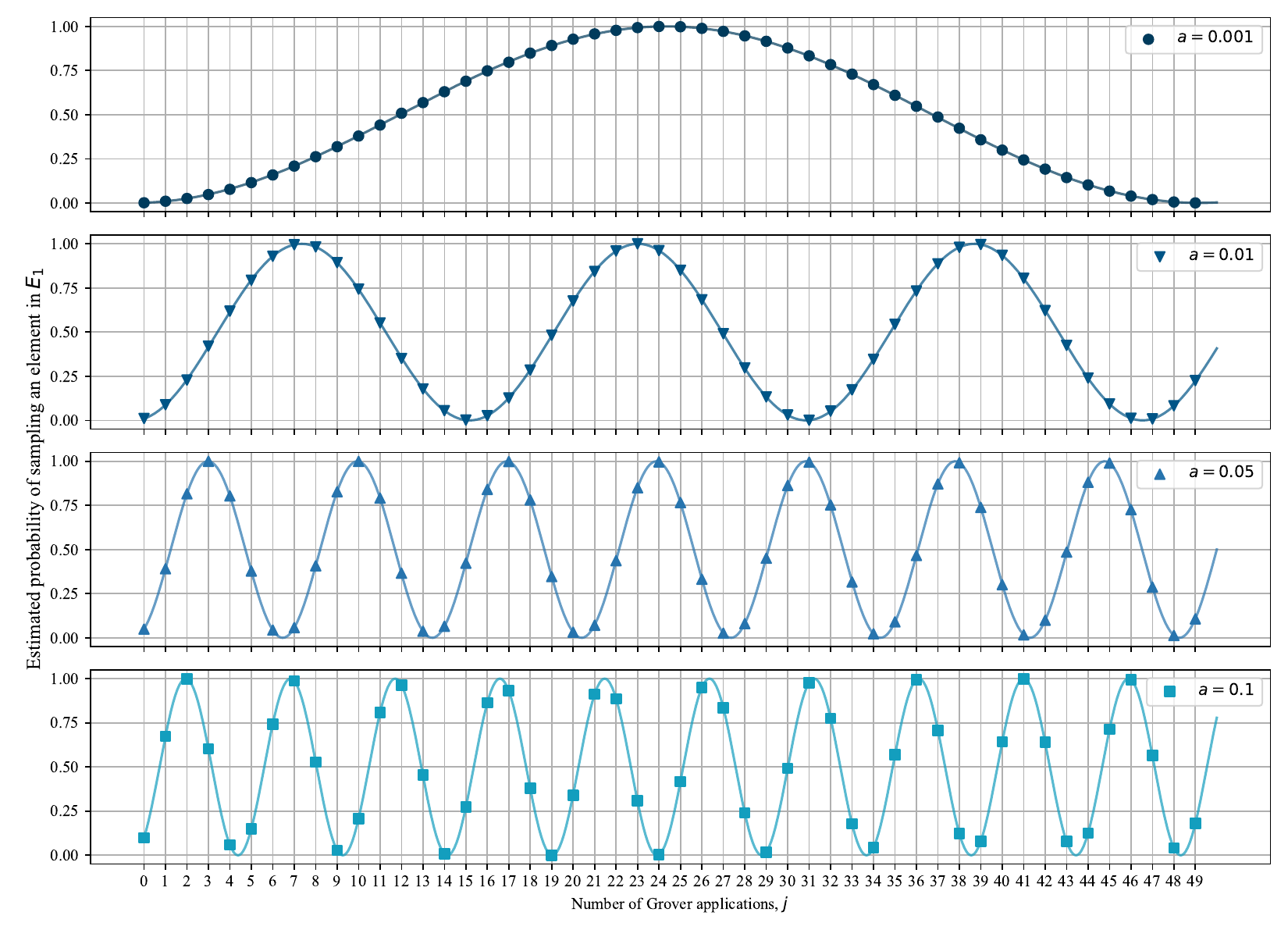}
	\caption{Probability of sampling an element from $E_1$ as a function of the number of Grover operations ($j$) applied to the quantum circuit, for different values of the initial sampling probability, $a$.}
	\label{fig:sec5_qaa}
\end{figure}

Figure \ref{fig:sec5_qaa} illustrates how the probability of sampling an element from $E_1$ varies for different values of $a$, the initial probability, and $j$, the number of Grover operations that are applied to the quantum circuit. Since we are interested in significantly improving the sampling probabilities of marked items, we limit ourselves to the case where sampling them initially is highly challenging, i.e., situations where $a < 0.1$. 

A series of important remarks can be observed from Figure \ref{fig:sec5_qaa}. First, note that while the figure shows continuous lines tracing the curve given by Equation \ref{eq:sec_5_3}, the number of Grover operations, $j$, must be an integer value, and therefore the achievable probabilities are those indicated by the markers in each plot. Second, as expected from Equation \ref{eq:sec_5_3}, the variation in $p(E_1)$ exhibits a sinusoidal pattern, with an oscillation period that decreases as the initial sampling probability, $a$, increases. This has two important consequences. First, for smaller values of the initial probability $a$, the highest is the number of Grover operations that are required to achieve a significant increase in the final sampling probability, $p(E_1)$. Second, it is possible to decrease the final sampling probability by applying an excessive number of Grover operations, and therefore selecting the ideal value for $j$ is an important practical consideration. Finally, all curves can reach values close to $1$. This is an outstanding result since it indicates that the QAA algorithm can, in principle, create a quantum circuit for which it is possible to exclusively obtain marked elements by the oracle function, completely ignoring those elements that belong to $E_0$. 

The QAA algorithm represents an outstanding result in quantum computation since it provides a feasible approach to selectively increase the likelihood of sampling marked outcomes through the definition of a suitable oracle function. The only condition for its use is that the original probability model is encoded as a quantum circuit, which is why it results fundamental to first translate the traditional Fault Tree model into its quantum equivalent, as shown in Section \ref{sec:qft}.

In the following section, this idea is first implemented in what we have termed as a \textit{Naive MCS identification approach}, with the objective of defining the quantum circuit forms of the unitary matrices $S_f$, $S_0$, and $A$ within the context of QFTs. In Section \ref{subsec:mcs}, we improve upon this \textit{naive} approach to generate the proposed algorithm for MCS identification.

\subsection{Naive approach for the identification of MCS using quantum computation.} \label{subsec:naive}

In this section, we present a naive approach that combines the QFT model and the QAA algorithm to increase the probability of measuring a state that represents an MCS. As we shall see, this increase is only marginal and therefore will be improved considerably in our proposed algorithm.

 Let us begin by translating a Fault Tree model into a QFT circuit, following Algorithm 1, by setting the failure probabilities of every basic event as $p_i=0.5,\  \forall i \in \{1,...,N_{BE}\}$. This modeling choice allows us to explore the outcome space uniformly, making every basic event configuration equally likely. The resulting quantum state is $\ket{\psi}=U_{FT} \ket{0}^{\otimes N}$. Then, let us prepare the Grover operator as $Q=A^{\dagger}S_0 A S_f$, with $A=U_{FT}$. For this, it is required to define the series of quantum gates and generate the matrices $S_f$ and $S_0$, as defined in Equations \ref{eq:sec_5_1} and \ref{eq:sec_5_2}, respectively.

For the construction of $S_f$, we define the oracle function as $f_{FT}: \{0,1\}^{N_{BE}}\rightarrow \{0,1\}$, which is a function that takes a given outcome of the basic events, and returns a $1$ if the system failed, and a $0$ if it did not. Note that this function is already implemented in the QFT model, with the output represented in the measurement of the $\ket{TOP}$ qubit. Consequently, as explained in Section \ref{subsec:U}, we can change the sign of all complex coefficients corresponding to states where $\ket{TOP}=\ket{1}$ by simply applying a Pauli-Z gate to the $\ket{TOP}$ qubit. This simple operation acts as an oracle, identifying all configurations where the system fails, i.e., the cut sets.

\begin{figure}[h]
	\centering
        \includegraphics[width=0.55\textwidth]{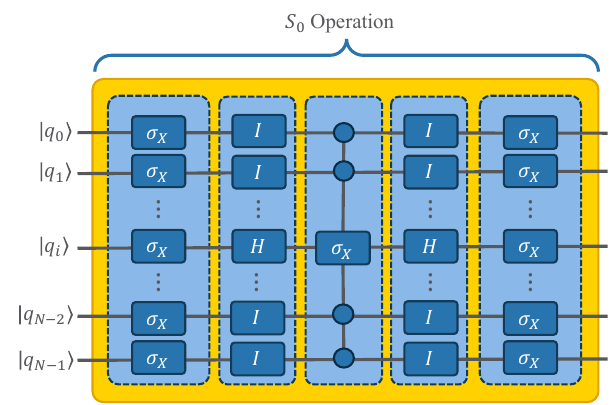}
	\caption{Quantum circuit representation of the $S_0$ operation. Note that the effect on the quantum state is the same independently of which qubit is selected as $\ket{q_i}$.}
	\label{fig:sec5_s0}
\end{figure}

For the $S_0$ operation, while the matrix shown in Equation \ref{eq:sec_5_2} is very simple, its implementation in a quantum circuit using the set of unitary gates shown in Section \ref{subsec:U} is not trivial. Indeed, to generate it, the following operations are applied to the qubit registry. First, a Pauli-X gate is applied to all qubits in the registry. Then, a qubit is selected from the registry and a Hadamard gate is applied over it. The third step is to apply a multi-CNOT gate over the registry, using as the “target” qubit the same qubit over which the Hadamard gate was applied. Finally, the operations are reversed, first by applying a second Hadamard gate over the selected qubit, and then by applying another round of Pauli-X gate over all qubits in the registry. The quantum circuit to produce the operation $S_0$ is summarized graphically in Figure \ref{fig:sec5_s0}. Note that it does not matter which qubit is selected from the registry to act as the “target” qubit in the operation; as long as the same qubit receives the Hadamard gates, the final result is the same. The overall effect of applying $S_0$ to a registry of length $N$ is to apply a negative sign to the complex coefficient that accompanies the canonical vector $\ket{e_0}$.

The successive application of the Grover operator $Q$ to the state prepared as $\ket{\psi}=U_{FT} \ket{0}^{\otimes N}$ tends to increase the likelihood of sampling the states marked by the $S_f$ operator. In this case, the effect is to increase the likelihood of sampling cut sets, i.e., states where $\ket{TOP}=\ket{1}$. As inferred from Equation \ref{eq:sec_5_3}, the variation in the sampling probability as the number of applications of $Q$ increases will not be monotonous; instead, it will vary in a sinusoidal pattern.

While this approach successfully increases the likelihood of identifying cut sets in the fault tree, it does not necessarily increase the likelihood of sampling MCS to the same extent. The reason for this is that in fault trees, MCS are in general much scarcer than cut sets and, therefore, the improvement provided by this \textit{naive} approach towards MCS identification could be negligible.

\begin{figure}[h]
\centering
\begin{subfigure}{0.6\textwidth}
    \includegraphics[width=\textwidth]{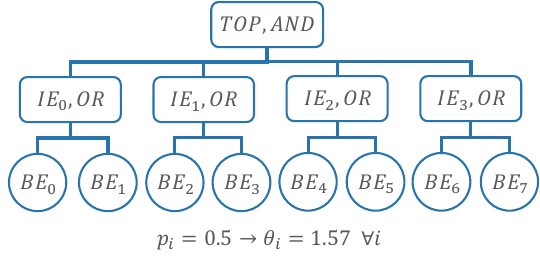}
    \caption{Fault Tree composed by $N_{BE}=8$ basic events and $N_{IE}=4$ intermediary events.}
\end{subfigure}
\hfill
\begin{subfigure}{0.9\textwidth}
    \includegraphics[width=\textwidth]{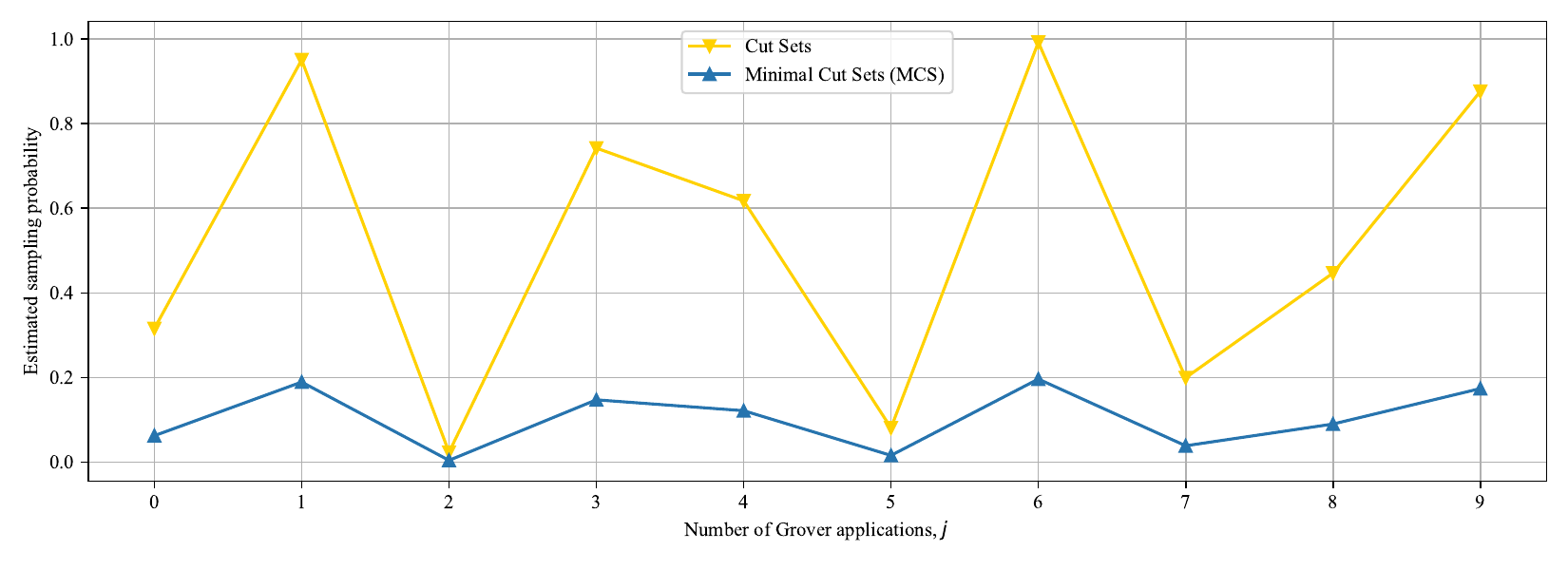}
    \caption{Variation of the cut set and minimal cut set estimated sampling probability for different numbers of Grover operators applied to the QFT model.}
\end{subfigure}
\caption{Numerical results corresponding to applying the \textit{naive} approach to a QFT generated from the Fault Tree model shown in (a). Note that from the $2^{N_{BE}}=256$ possible outcomes, $81$ correspond to cut sets, while only $16$ correspond to minimal cut sets (MCS).}
\label{fig:sec5_ft}
\end{figure}

To exemplify this effect, we make use of the Fault Tree presented in Figure \ref{fig:sec5_ft} (a). Figure \ref{fig:sec5_ft} (b) depicts the variation of the estimated probability of sampling configurations that represent cut sets (yellow line) and minimal cut sets (blue line) as the number of Grover operations increases. The case where $j=0$ represents sampling from the original fault tree and can be considered equivalent to performing traditional Monte Carlo. Note how the probability of sampling a cut set increases until $j=1$, then decreases, and then the cycle is approximately repeated in a sinusoidal pattern, as expected. The probability of sampling a minimal cut set follows a similar pattern. However, it is notoriously lower due to the abundance of cut sets as compared to MCS in this fault tree.

The proposed algorithm improves upon this situation expanding the QFT circuit to identify MCS instead of cut sets, and thus targeting these states directly with the QAA algorithm. The remainder of this section is devoted to describing the proposed methodology.

\subsection{Proposed Algorithm for the identification of Minimal Cut Sets using Quantum Computation} \label{subsec:mcs}

In Section \ref{subsec:qaa} we show how the QAA algorithm can be used to increase the likelihood of a group of marked states when measuring quantum states. In Section \ref{subsec:naive}, we use that concept to increase the likelihood of sampling configurations that cause the system to fail, by identifying the marked states as those where the TOP event occurs, in an attempt to increase the likelihood of measuring minimal cut sets (MCS). However, as shown in Figure \ref{fig:sec5_ft}, this \textit{naive} approach has several drawbacks and therefore needs to be improved. In this section, we show an improved alternative to MCS identification using quantum computation.

The overarching idea behind our proposed algorithm is to modify the approach shown in Section \ref{subsec:naive}, by generating a quantum circuit that encodes a Boolean function that recognizes whether a basic event configuration is an MCS or not. We will denote this function as $f_{MCS}:\{0,1\}^{N_{BE}} \rightarrow \{0,1\}$. To create this function, we make use of the description of an MCS presented in Definition 2, Section \ref{sec:ft}. We present the definition here as well for the reader’s convenience: a configuration of basic events $\vec{x}_{BE}$ is a MCS if and only if it fulfills two conditions: first, it makes the system fail (it is a cut set), and second, preventing the failure of any of its failed basic events would also prevent the failure of the system (it is irreducible). Equation \ref{eq:sec_5_4} uses this definition to generate $f_{MCS}$, where the function $s:\{0,1\}{N_{BE}} \rightarrow {0,1}^{N_{BE}}$ is a \textit{turning off} function that sets the i-th component of $\vec{x}_{BE}$ to $0$, while maintaining all other components equal. The set $F$ is defined as $F=\{i | x_{BE_i}=1\}$, i.e., a collection of basic events indexes that failed in configuration $\vec{x}_{BE}$.

\begin{equation} \label{eq:sec_5_4}
    f_{MCS}(\vec{x}_{BE}) = f_{FT}(\vec{x}_{BE}) \land \bigwedge_{i\in F} \neg f_{FT}(s(\vec{x}_{BE},i))
\end{equation}

Note that the first term corresponds to the outcome of the original fault tree under configuration $\vec{x}_{BE}$, while the following $|F|$ logical clauses correspond to the negated state of the system under a modified configuration where the basic event $i \in F$ does not fail. 

To implement $f_{MCS}$ into a quantum circuit, we will use both the QFT framework presented in Section \ref{sec:qft} and the reversibility of quantum operations, described in Section \ref{subsec:U}. The quantum circuit implementation of the MCS Boolean function, which we will denote $U_{MCS}$, starts by initializing a registry of length $N=2N_{BE}+N_{IE}+3$ qubits. We divide this registry into the following sub-registries (ordered from the top to the bottom in the quantum circuit):

\begin{itemize}
    \item $\kbe$ for representing the basic events.
    \item $\kie$ for representing intermediary events.
    \item $\ket{TOP}$ for representing the outcome of the TOP event under the original fault tree.
    \item $\ktop$ for representing the negated outcome of the TOP event under a modified fault tree, where the i-th basic event is forced to not fail.
    \item $\ket{MCS}$ to represent whether a particular measurement outcome represents an MCS or not.
    \item $\ket{aux}$, prepared in the $\ket{0}$ state to use as a replacement for each one of the basic event’s qubits in the algorithm.
\end{itemize}

After initializing all the qubits in the registry in the $\ket{0}$ state, the QFT circuit is applied over the registries $[ \kbe, \kie, \ket{TOP}]$, followed by applying the adjoint operation $U_{IE}^{\dagger}$ over the registry $\kie$. The result of this operation is a quantum circuit that correctly encodes the first operator from Equation \ref{eq:sec_5_4} into the $\ket{TOP}$ qubit while returning the qubits $\kie$ to their original state for future computation.

To implement the rest of the operator, a series of $N_{BE}$ unitary operations are applied to the quantum circuit. The objective of each of these stages is to encode into $\ket{TOP_i}, i \in \{1,...,N_{BE}\}$, a $\ket{1}$ if the system does not fail under the restriction $\ket{BE_i}=\ket{0}$, and a $\ket{0}$ otherwise. To do this, we apply the unitary operation described in Equation \ref{eq:sec5_5} to  to the quantum registry. 

\begin{equation} \label{eq:sec5_5}
    U_i=U_{IE}(\ket{BE_i})\ U_{TOP}(\ket{TOP_i})\ CNOT(\ket{BE_i},\ket{TOP_i})\ U_{IE}^{\dagger}(\ket{BE_i})
\end{equation}

Let us unpack and explain each of the pieces that compose this operation:

\begin{itemize}
    \item $U_{IE}(\ket{BE_i})$ is an analogous operation as $U_{IE}$, but replacing the input qubit $\ket{BE_i}$ for $\ket{aux}$, forcing $\ket{BE_i}$ to not fail.
    \item $U_{TOP}(\ket{TOP_i})$ is analogous to the operation $U_{TOP}$, but storing the system’s outcome under the modified configuration $s(\vec{x}_{BE},i)$ in the $\ket{TOP_i}$ qubit.
    \item $CNOT(\ket{BE_i},\ket{TOP_i})$ is a Control-NOT operation taking as a control qubit $\ket{BE_i}$, and as the target qubit $\ket{TOP_i}$. This operation is fundamental since it allows the selective negation of the fault tree outcome only for those cases where $\ket{BE_i}$ is measured as $\ket{1}$, i.e., $i \in F$ in Equation \ref{eq:sec_5_4}.
    \item The final operation, $U_{IE}^{\dagger}(\ket{BE_i})$ is the adjoint of $U_{IE}(\ket{BE_i})$. By applying it, we again free the qubits in the registry $\kie$ to allow for the verification of a new clause in the algorithm.
\end{itemize}

Once the operators $\{U_i\}_{i=1}^{N_{BE}}$ have been applied, an MCS can be identified by using the set $\{\ket{TOP}, \ktop \}$ as the input for a quantum AND gate and storing the result in the qubit $\ket{MCS}$. This finalizes the implementation of Equation \ref{eq:sec_5_4} in a quantum circuit. Algorithm \ref{Algo2} summarizes this procedure.

\begin{algorithm}[h]
\kwInputs{\\ \begin{itemize}
 \item Inputs to generate QFT:
 \begin{itemize}
      \item Set of basic events $\be$, where each basic event is associated with a probability of failure $p_i$.
    \item Set of intermediary events $\ie$, where each intermediary event is associated with a generating logical operator $O_i\in\{AND,OR\}$, and a set of inputs $S_i$. The set of intermediary events is ordered in such a way that $IE_i$ is not an input of $IE_j$ if $i\leq j$.
    \item A TOP event, described by a generating logical operator $O_{TOP} \in \{AND,OR\}$, and a set of inputs $S_{TOP}$.
 \end{itemize}

\end{itemize}}

\KwResult{A unitary operation $U_{MCS}$ that encodes the function $f_{MCS}$}
\textbf{Algorithm:}
\begin{enumerate}
\item Preparation of qubit registries:

        \begin{itemize}
            \item $\kbe$ for representing the basic events.
            \item $\kie$ for representing intermediary events.
            \item $\ket{TOP}$ for representing the outcome of the TOP event under the original fault tree.
            \item $\ktop$ for representing the negated outcome of the TOP event under a modified fault tree, where the i-th basic event is forced to not fail.
            \item $\ket{MCS}$ to represent whether a particular measurement outcome represents an MCS or not.
            \item $\ket{aux}$, prepared in the $\ket{0}$ state to use as a replacement for each one of the basic event’s qubits in the algorithm.
        \end{itemize}

\item 	Prepare QFT model (Algorithm \ref{Algo1}), using qubit registries $[ \kbe, \kie, \ket{TOP}]$.

\item Apply $U_{IE}^{\dagger}$ to qubit registry $\kie$.

\item For $i \in \{1,...,N_{BE}\}$:
        \begin{itemize}
            \item If $O_{TOP} = AND$:
            \begin{itemize}
                \item 	Apply $U_i=U_{IE}(\ket{BE_i})\ U_{TOP}(\ket{TOP_i})\ CNOT(\ket{BE_i},\ket{TOP_i})\ U_{IE}^{\dagger}(\ket{BE_i})$ to the quantum registry.
            \end{itemize}
        \end{itemize}

\item Apply a CNOT gate (Quantum AND) with the registry $\{ \ket{TOP}, \ktop \}$ as control qubits and $\ket{MCS}$ as target qubit.
\end{enumerate}
 \caption{Encoding of $f_{MCS}$ into a quantum circuit.}
 \label{Algo2}
\end{algorithm}

With the implementation of $f_{MCS}$ as the unitary operation $U_{MCS}$, the QAA algorithm can be applied to the resulting quantum circuit to increase the likelihood of sampling minimal cut sets. For this, we define $A=U_{MCS}$ and $S_f$ as a single Pauli-Z gate in the $\ket{MCS}$ qubit. $S_0$ has the same definition as the one described in Section \ref{subsec:qaa}.  Our proposed algorithm is formalized in Algorithm \ref{Algo3}.

\begin{algorithm}[h]
\kwInputs{\\ \begin{itemize}
 \item Inputs to generate QFT:
 \begin{itemize}
      \item Set of basic events $\be$, where each basic event is associated with a probability of failure $p_i$.
    \item Set of intermediary events $\ie$, where each intermediary event is associated with a generating logical operator $O_i\in\{AND,OR\}$, and a set of inputs $S_i$. The set of intermediary events is ordered in such a way that $IE_i$ is not an input of $IE_j$ if $i\leq j$.
    \item A TOP event, described by a generating logical operator $O_{TOP} \in \{AND,OR\}$, and a set of inputs $S_{TOP}$.
 \end{itemize}
\item $J$, number of applications of the Grover operator.
\end{itemize}}

\KwResult{$K$ samples of the system’s configurations. It is expected that a considerable portion of these samples correspond to MCS. }
\textbf{Algorithm:}
\begin{enumerate}
\item For $k \in \{1,...,K\}$:
\begin{itemize}
    \item 	Prepare quantum circuit for $U_{MCS}$,  Algorithm \ref{Algo2}.
    \item For $j \in \{1,...,J\}$:
    \begin{itemize}
        \item 	Apply to the quantum circuit $Q=U_{MCS}^{\dagger}S_0 U_{MCS} S_f$.
    \end{itemize}
    \item Measure all qubits and store $\vec{x}$ as sample $k$.
\end{itemize}
\end{enumerate}
 \caption{Minimal Cut Set identification using Quantum Computation.}
 \label{Algo3}
\end{algorithm}

In the following section, we test the proposed algorithm to assess its capabilities and how it compares to a standard Monte Carlo sampling approach.

\section{Experimental Validation} \label{sec:exp}

\subsection{Experimental Setup} \label{subsec:exp_setup}

As of 2024, only small-capacity and error-prone quantum computers are available for researchers and practitioners. While an exciting tool, current quantum hardware is still the target of intense exploration and development. Thus, these devices are not yet sufficient to carry out large computations without errors contaminating the delicate quantum state that they generate and consequently diminishing the reliability of the computation.

For that reason, all the experiments in this paper are executed on a quantum simulator, which is a software stacks that run on a traditional computer and is designed to simulate the behavior of an error-corrected, noise-free quantum computer. The use of a quantum simulator has the advantage of enabling the execution of quantum approaches in ideal circumstances to study their algorithmic effectiveness. 

However, a notable disadvantage of quantum simulators is the relatively small size of the quantum states they can represent. Recall that the size of quantum states and gates grows exponentially with the number of qubits in the operation. As such, a traditional computer executing a quantum simulator has a maximum capacity (measured in the length of the qubit registry it can represent) dictated by its RAM memory. 

In the case of this paper, the workstation used is equipped with a 32-core CPU and 128GB of RAM, and therefore the qubit registry was limited to 23 qubits. For this reason, the Fault Tree examples shown in this section are designed to portray the capabilities of the algorithm, instead of bench-marking it by using large Fault Trees against state-of-the-art approaches designed to run in traditional computers.

\subsection{Theoretical Comparison with Monte Carlo Sampling}\label{subsec:MC}

We start the proposed algorithm validation with a theoretical comparison against traditional Monte Carlo sampling in the task of identifying all MCS in a given Fault Tree.

For this, recall that a consequence of assuming the failure probabilities to be $p_i=0.5, \forall i \in \{1,...,N_{BE}\}$  is that all basic event configurations are equally likely, and therefore the probability of finding an MCS via Monte Carlo sampling is equal to $p_{MC}=N_{MCS}/2^{N_{BE}}$, where $N_{MCS}$ is the number of MCS configurations and $N_{BE}$ is the number of basic events in the Fault Tree, respectively. Using this probability, the expected number of samples required to find all MCS can be computed by adapting the well-known \textit{Coupon Collector’s Problem} \cite{neal2008generalised} to the case where only some of the coupons are of interest to the collector. We start by defining the random variable $X_k$ as the number of samples required until the k-th MCS is found, assuming no order in particular. If we define a “successful” experiment as obtaining a sample that corresponds to an MCS, then $X_k$ is a geometric random variable with “success” probability equal to $p_k=(N_{MCS}-k+1)/2^{N_{BE}}$  and expected value $\mathbb{E}[X_k]=1/p_k$. Then, the total expected number of samples using Monte Carlo sampling, $\mathbb{E}[X_{MC}]$, can then be computed as:

\begin{equation} \label{eq:6_1}
\begin{split}
\mathbb{E}[X_{MC}] & = \sum_{k=1}^{N_{MCS}} \mathbb{E}[X_k] = \mathbb{E} \left[ \sum_{k=1}^{N_{MCS}} X_k \right] = \mathbb{E} \left[ \sum_{k=1}^{N_{MCS}} \frac{1}{p_k} \right] \\
& = 2^{N_{BE}} \cdot \left(\frac{1}{N_{MCS}} + \frac{1}{N_{MCS}-1} + \dots + 1 \right) \\
& = 2^{N_{BE}} H_n(N_{MCS})\\
\end{split}
\end{equation}

where $H_n(N_{MCS})$ is the n-th harmonic number computed at $n=N_{MCS}$.

On the other hand, the effect of the proposed algorithm is to increase the base probability of sampling an MCS in accordance with Equation \ref{eq:6_2}, adapted here from Equation \ref{eq:sec_5_3} to fit the context of Fault Trees:

\begin{equation} \label{eq:6_2}
    p_{QAA}(j) = \sin ^2 \left((2j+1)\ \mathrm{asin}(\sqrt{p_{MC}})\right) 
\end{equation}

where $j$ is the number of Grover operations that are applied to the quantum model, as explained in Section \ref{subsec:mcs}. Consequently, by updating the \textit{success} probability of each random variable $X_k$ to $p_k=p_{QAA}(j)(N_{MCS}-k+1)/N_{MCS}$, we can use the same approach to compute the expected number of samples required by the proposed algorithm as:

\begin{equation} \label{eq:6_3}
    \mathbb{E}[X_{QAA}] = \sum_{k=1}^{N_{MCS}} \mathbb{E}[X_k] = \frac{N_{MCS}}{p_{QAA}(j)} \cdot H_n(N_{MCS})
\end{equation}

To compare the expected number of samples required by both approaches, a ratio is computed in Equation \ref{eq:6_4}:

\begin{equation} \label{eq:6_4}
    r=\frac{\mathbb{E}[X_{MC}]}{\mathbb{E}[X_{QAA}]} = \frac{p_{QAA}(j) \cdot 2^{N_{BE}}}{N_{MCS}}
\end{equation}

Equation \ref{eq:6_4} shows that if the proposed algorithm is able to produce values of $p_{QAA}(j)$ that are close to $1$, then traditional Monte Carlo sampling will require a number of samples that is $\sim 2^{N_{BE}}/N_{MCS}$  times higher than what is required by using the proposed algorithm. While the number of MCS in a fault tree will depend on its structure, it is in general expected to be orders of magnitude lower than the number of possible configurations. Consequently, the proposed algorithm has the potential to considerably improve on the complex task of identifying all MCS in a given Fault Tree.

\subsection{Numerical comparison with the quantum-based Naive approach} \label{subsec:final}

Section \ref{subsec:MC} shows that the proposed algorithm presents a considerable theoretical benefit with respect to traditional Monte Carlo sampling. However, how does it compare with the naive approach described in Section \ref{subsec:naive}? In other words, it is worth it to use a higher number of qubits to identify MCS instead of using the occurrence of the TOP event as a proxy for the oracle operator?

To answer these questions, we utilize the fault tree introduced in Figure 6 (a) as the basis to implement the proposed algorithm for MCS identification. Recall that the fault tree shown in Figure \ref{fig:sec5_ft} (a) has $16$ out of $256$ configurations that represent minimal cut sets.

\begin{figure}[h]
	\centering
        \includegraphics[width=0.95\textwidth]{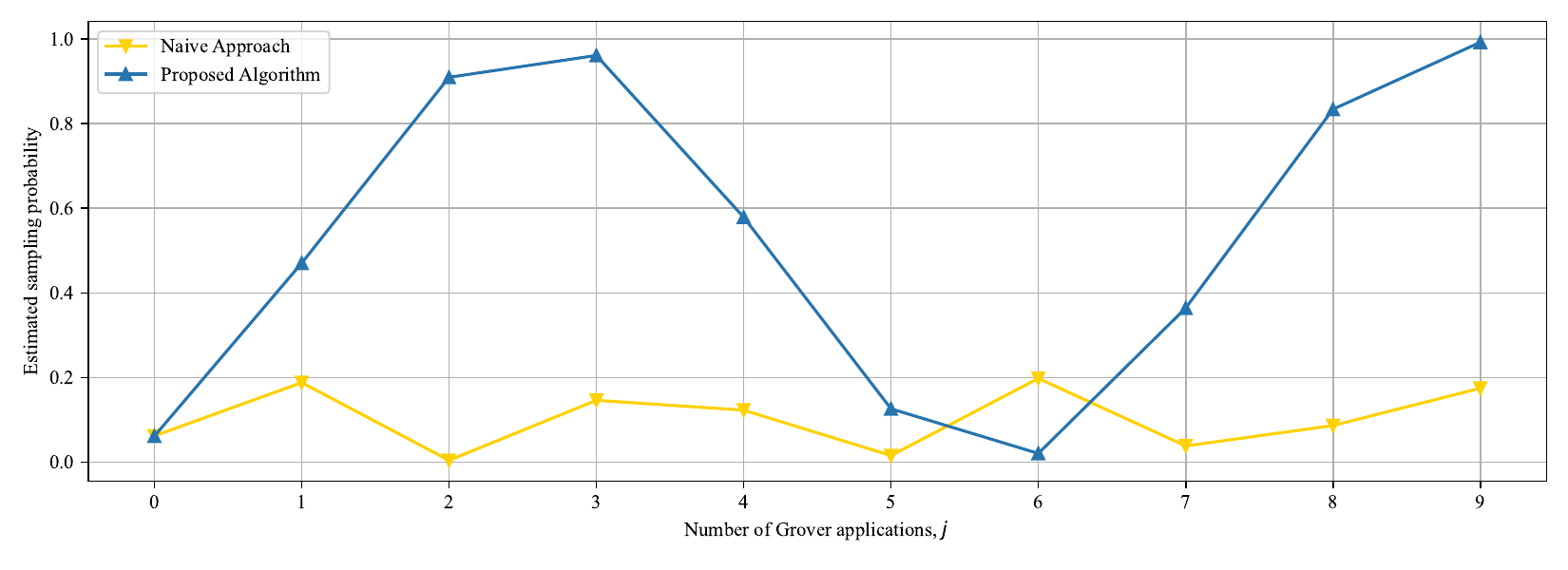}
	\caption{Numerical comparison between the proposed algorithm and the naive approach in terms of the estimated MCS sampling probability, $p(MCS)$.}
	\label{fig:sec6}
\end{figure}

Figure \ref{fig:sec6} shows, for both the proposed algorithm and the naive approach, the estimated probability of sampling an MCS configuration, $p(MCS)$, as a function of the number of Grover operator applications, $j$. In each case, a total of $10^5$ samples were obtained from the measurement operations to accurately estimate these probabilities.

The effect of using a quantum circuit that implements $f_{mcs}$ (proposed algorithm) instead of $f_{top}$ (naive approach) to generate the oracle operator can be clearly seen in Figure \ref{fig:sec6}, where considerably higher sampling probabilities are achieved by the former compared to the latter. In particular, for the case $j=3$ and $j=9$, the proposed algorithm is capable of achieving values for $p(MCS)$ close to $1$, which indicates that all samples retrieved in these cases correspond to MCS.

Figure \ref{fig:sec6} also allows the numerical verification of the theoretical discussion presented in Section \ref{subsec:MC}. For this particular fault tree, the probability of sampling a minimal cut set using Monte Carlo sampling is $p_{MC}=N_{MCS}/2^{N_{BE}}=16/256= 0.0625$, which coincides with the estimated value of both the naive approach and proposed algorithm for the case {j=0}. This result is expected since the QAA is not applied to increase the sampling probabilities. The naive approach obtains, in the best case corresponding to {j=6}, an enhanced sampling probability of $p_{QAA}^{naive}=0.196$ , while the proposed algorithm achieves a maximum sampling probability of $p_{QAA}^{proposed}=0.992$ for the case {j=9}. Using these probabilities, the expected number of samples required by each technique is shown in Table \ref{tab:sec6}.

\begin{table}[h]
\centering
\caption{Expected number of samples $\mathbb{E}[X]$ for each technique.}
\label{tab:sec6}
\begin{tabular}{lll}\toprule

Monte Carlo Sampling & Naive Approach (Section \ref{subsec:naive}) & Proposed Algorithm (Section \ref{subsec:mcs})  \\ \midrule \addlinespace[1.5ex]
$\mathbb{E}[X_{MC}]=865$ & $\mathbb{E}[X_{QAA}^{naive}]=276$ & $\mathbb{E}[X_{QAA}]=55$ \\ \addlinespace[1.5ex] 
\bottomrule
\end{tabular}
\end{table}

Note how the proposed algorithm requires a much lower number of samples to find all MCS in the fault tree than the Monte Carlo sampling. Additionally, the proposed algorithm is considerably better than the naïve approach as well, requiring around five times fewer samples.

\section{Concluding Remarks} \label{sec:concluding}

This paper proposes a novel algorithm for the identification of Minimal Cut Sets in Fault Trees using quantum computation. For this, the traditional fault tree model is translated into a quantum circuit equivalent, preserving all the logical and probabilistic relationships between components. The quantum fault tree model is combined with the Quantum Amplitude Amplification algorithm to generate a novel algorithm capable of increasing the sampling probability of states that represent Minimal Cut Sets. The proposed algorithm was theoretically and numerically compared against both quantum based and traditional Monte Carlo approaches.

The results indicate that the proposed algorithm has the potential to increase multiple times the probability of randomly sampling a system configuration that corresponds to a minimal cut set, thus reducing in orders of magnitude the amount of computation required to find all of them. However, the proposed approach contains a series of limitations that require further consideration and constitute ideal paths for future research. First, the identification of the optimal number of Grover operators, $j$, to maximize the minimal cut set sampling probability is dependent on the initial sampling probability, which is a priori unknown in a practical, real-life system modeled via a fault tree. Without knowledge about this quantity, the algorithm must undergo a stage of testing different values of this parameter until finding one that produces a suitable result, possibly increasing the time complexity of the algorithm. Identifying approaches to estimate this quantity beforehand at a low algorithmic cost could potentially alleviate this limitation. Second, when compared to an alternative quantum approach, which we have denominated here as the “naive approach”, the proposed algorithm uses a larger number of qubits. This could make its practical application challenging in contemporary noisy quantum devices. While the proposed algorithm remains linear in the number of qubits, identifying ways to decrease the qubit count could increase the applicability of the proposed model, especially for its implementation in early quantum computers. Third, the proposed algorithm is only applicable to coherent Fault Trees. While a common limitation in the literature, imposing this condition also restricts the applicability of our quantum approach to a non-negligible portion of systems in practice. Finally, the failure of basic events is modeled in this paper as Bernoulli probability distributions. While a common approach in Fault Tree Analysis, the inclusion of other distributions commonly used in Probabilistic Risk Analysis, such as the Weibull distribution will be required to extend the proposed algorithm towards non-standard Fault Trees, such as Dynamic or Fuzzy models.

Overall, the approach proposed in this paper constitutes an important step towards integrating quantum computation into traditional models used in the Probabilistic Risk Assessment of complex engineering systems, enabling the generation of highly efficient protocols to perform fundamental PRA and reliability tasks.

\bibliographystyle{ieeetr}
\bibliography{template}  

\begin{thebibliography}{10}

\bibitem{watson1961launch}
H.~A. Watson {\em et~al.}, ``Launch control safety study,'' {\em Bell labs}, 1961.

\bibitem{fussell1976fault}
J.~Fussell, ``Fault tree analysis: concepts and techniques,'' in {\em Pressure vessels and piping: design and analysis. IV}, 1976.

\bibitem{wang2016fault}
J.~Wang, F.~Wang, S.~Chen, J.~Wang, L.~Hu, Y.~Yin, and Y.~Wu, ``Fault-tree-based instantaneous risk computing core in nuclear power plant risk monitor,'' {\em Annals of Nuclear Energy}, vol.~95, pp.~35--41, 2016.

\bibitem{volkanovski2009application}
A.~Volkanovski, M.~{\v{C}}epin, and B.~Mavko, ``Application of the fault tree analysis for assessment of power system reliability,'' {\em Reliability Engineering \& System Safety}, vol.~94, no.~6, pp.~1116--1127, 2009.

\bibitem{cheng2013application}
C.-Y. Cheng, S.-F. Li, S.-J. Chu, C.-Y. Yeh, and R.~J. Simmons, ``Application of fault tree analysis to assess inventory risk: a practical case from aerospace manufacturing,'' {\em International Journal of Production Research}, vol.~51, no.~21, pp.~6499--6514, 2013.

\bibitem{rogith2017using}
D.~Rogith, M.~S. Iyengar, and H.~Singh, ``Using fault trees to advance understanding of diagnostic errors,'' {\em The Joint Commission Journal on Quality and Patient Safety}, vol.~43, no.~11, pp.~598--605, 2017.

\bibitem{baek2021application}
S.~Baek and G.~Heo, ``Application of dynamic fault tree analysis to prioritize electric power systems in nuclear power plants,'' {\em Energies}, vol.~14, no.~14, p.~4119, 2021.

\bibitem{kabir2020hybrid}
S.~Kabir, K.~Aslansefat, I.~Sorokos, Y.~Papadopoulos, and S.~Konur, ``A hybrid modular approach for dynamic fault tree analysis,'' {\em IEEE Access}, vol.~8, pp.~97175--97188, 2020.

\bibitem{mahmood2013fuzzy}
Y.~A. Mahmood, A.~Ahmadi, A.~K. Verma, A.~Srividya, and U.~Kumar, ``Fuzzy fault tree analysis: a review of concept and application,'' {\em International Journal of System Assurance Engineering and Management}, vol.~4, pp.~19--32, 2013.

\bibitem{vatn1992finding}
J.~Vatn, ``Finding minimal cut sets in a fault tree,'' {\em Reliability Engineering \& System Safety}, vol.~36, no.~1, pp.~59--62, 1992.

\bibitem{nielsen2010quantum}
M.~A. Nielsen and I.~L. Chuang, {\em Quantum computation and quantum information}.
\newblock Cambridge university press, 2010.

\bibitem{brassard2002quantum}
G.~Brassard, P.~Hoyer, M.~Mosca, and A.~Tapp, ``Quantum amplitude amplification and estimation,'' {\em Contemporary Mathematics}, vol.~305, pp.~53--74, 2002.

\bibitem{ruijters2015fault}
E.~Ruijters and M.~Stoelinga, ``Fault tree analysis: A survey of the state-of-the-art in modeling, analysis and tools,'' {\em Computer science review}, vol.~15, pp.~29--62, 2015.

\bibitem{lee1985fault}
W.-S. Lee, D.~L. Grosh, F.~A. Tillman, and C.~H. Lie, ``Fault tree analysis, methods, and applications: a review,'' {\em IEEE transactions on reliability}, vol.~34, no.~3, pp.~194--203, 1985.

\bibitem{vesely1981fault}
W.~E. Vesely, F.~F. Goldberg, N.~H. Roberts, D.~F. Haasl, {\em et~al.}, {\em Fault tree handbook}.
\newblock Systems and Reliability Research, Office of Nuclear Regulatory Research, US, 1981.

\bibitem{rauzy1993new}
A.~Rauzy, ``New algorithms for fault trees analysis,'' {\em Reliability Engineering \& System Safety}, vol.~40, no.~3, pp.~203--211, 1993.

\bibitem{coudert1993fault}
O.~Coudert and J.~C. Madre, ``Fault tree analysis: 10/sup 20/prime implicants and beyond,'' in {\em Annual Reliability and Maintainability Symposium 1993 Proceedings}, pp.~240--245, IEEE, 1993.

\bibitem{vesely1970prep}
W.~Vesely and R.~Narum, {\em PREP and KITT: computer codes for the automatic evaluation of a fault tree}, vol.~1349.
\newblock US Atomic Energy Commission, 1970.

\bibitem{silva2022quantum}
G.~S.~M. Silva, T.~Parhizkar, and E.~L. Droguett, ``Quantum fault trees,'' {\em arXiv preprint arXiv:2204.10877}, 2022.

\bibitem{brand_d6_2022}
S.~Brand, A.~Laarman, and V.~Moret, ``D6. 6: {Divide} and quantum open source software,'' 2022.

\bibitem{grover1996fast}
L.~K. Grover, ``A fast quantum mechanical algorithm for database search,'' in {\em Proceedings of the twenty-eighth annual ACM symposium on Theory of computing}, pp.~212--219, 1996.

\bibitem{neal2008generalised}
P.~Neal, ``The generalised coupon collector problem,'' {\em Journal of Applied Probability}, vol.~45, no.~3, pp.~621--629, 2008.

\end{thebibliography}

\end{document}